\documentclass[10pt, conference, compsocconfz]{IEEEtran}

\usepackage{graphicx}

\usepackage{algorithm}
\usepackage{algorithmic}

\newcommand{\gid}{\textsc{GID}}
\newtheorem{example}{Example}[section]

\newcommand{\RN}[1]{%
  \textup{\uppercase\expandafter{\romannumeral#1}}%
}

\newcommand{\algorithmicbreak}{\textbf{break}}
\newcommand{\BREAK}{\STATE \algorithmicbreak}

\newcommand{\nop}[1]{}

\begin{document}






%

\title{{\em GID}: {\em G}raph-based {\em I}ntrusion {\em D}etection on Massive Process Traces for Enterprise Security Systems}

\author{
\IEEEauthorblockN{Boxiang Dong\IEEEauthorrefmark{1}, Zhengzhang Chen\IEEEauthorrefmark{2}, Hui Wang\IEEEauthorrefmark{1}, Lu-An Tang\IEEEauthorrefmark{2}, Kai Zhang\IEEEauthorrefmark{2}, Ying Lin\IEEEauthorrefmark{3}, \\ Wei Cheng\IEEEauthorrefmark{2}, Haifeng Chen\IEEEauthorrefmark{2} and Guofei Jiang\IEEEauthorrefmark{2}}
\IEEEauthorrefmark{1}Stevens Institute of Technology, Hoboken NJ\\
\IEEEauthorrefmark{2}NEC Laboratories America, Princeton NJ\\
\IEEEauthorrefmark{3}University of Washington, Seattle WA
}

\maketitle

\begin{abstract}
Intrusion detection system (IDS) is an important part of enterprise security system architecture. In particular, anomaly-based IDS has been widely applied to detect abnormal process behaviors that deviate from the majority. However, such abnormal behavior usually consists of a series of low-level heterogeneous events. The gap between the low-level events and the high-level abnormal behaviors makes it hard to infer which single events are related to the real abnormal activities, especially considering that there are massive ``noisy'' low-level events happening in between. Hence, the existing work that focus on detecting single entities/events can hardly achieve high detection accuracy. Different from previous work, we design and implement \gid, an efficient graph-based intrusion detection technique that can identify abnormal event sequences from a massive heterogeneous process traces with high accuracy. \gid\ first builds a compact graph structure to capture the interactions between different system entities. The suspiciousness or anomaly score 
of process paths is then measured by leveraging random walk technique to the constructed acyclic directed graph. To eliminate the score bias from the path length, the Box-Cox power transformation based approach is introduced to normalize
the anomaly scores so that the scores of paths of different lengths have the same distribution. 
The efficiency of suspicious path discovery is further improved by the proposed optimization scheme. 
We fully implement our \gid\ algorithm and deploy it into a real enterprise security system, 
and it greatly helps detect the advanced threats, and optimize the incident response. Executing \gid\ on system monitoring datasets showing that \gid\ is efficient (about 2 million records per minute) and accurate (higher than 80\% in terms of detection rate).   

\end{abstract}

%
%


%
%

%
%




\section{Introduction}
With computers and networked systems playing indispensable roles in almost every aspect of modern world of society such as industry, government, and economy, cyber-security undoubtedly bears the utmost importance in preserving right social orders. However, nowadays, serious cyber-attacks still keep happening, which causes significant financial loss and public tensions. One example is the leakage of sensitive, high-profile information from giant marketing establishments or financial institutions. According to a recent study by Ponemon Institute and IBM~\cite{ponemon2014cost},
data breaches cost companies an average of \$201 per record in 2014, and the total cost paid by organizations reaches \$5.9 millions.

To guarantee the information security in the network of computers, an intrusion detection system (IDS) is needed to keep track of the running status of the entire network and identify scenarios that are associated with potential attacks or malicious behaviours. 
Compared to signature-based methods, which can only detect attacks for which a signature has previously been created, anomaly-based intrusion detection aims at identifying unusual entities, events, or observations from a running system that deviate from its normal pattern of behaviours.
Detected anomaly patterns can be translated into critical actionable information that can significantly facilitate human decision making and mitigate the damage of cyber-attacks. 
Towards this end, anomaly-based intrusion detection, with a variety of algorithms proposed in recent years (e.g., \cite{lin2012intelligent,jyothsna2011review}), turns out to be a particularly useful tool.

However, building an efficient and accurate anomaly-based intrusion detection system is still challenging due to the nature of the system monitoring data. 
However, building an efficient and accurate anomaly-based intrusion detection system is still challenging due to the nature of the system monitoring data. 
First, IDS typically deals with a large volume of system event data (normally more than $10,000$ events per host per second) to detect possible malicious activities. 
Second, the variety of system entity types may necessitate high-dimensional features in subsequent processing. 
Such enormous 
feature space could easily lead to the problem, coined by Bellman as ``the curse of dimensionality'' \cite{Bellman1961}.  

More importantly, IDS often relies on a \textit{coordinated or sequential, not independent},
action of several system events to determine what state a given system is in.  The system monitoring data are typically low-level 
interactions between various entities (e.g., a program opens a file or connects to a server) with exact time stamps, while abnormal behaviors such as attempted intrusions are higher-level activities which usually involve multiple different events. For example, a network attack called Advanced Persistent Threat (APT)  is composed of a set of stealthy and continuous computer hacking processes, by first attempting to gain a foothold in the environment, then using the compromised systems as the access into the target network, followed by deploying additional tools that help fulfill the attack objective. 
The gap between the low-level events (e.g., process events) and the high-level abnormal behaviors (e.g., APT attacks) makes it hard to infer which events are related to the real abnormal activities, especially considering that there are massive ``noisy'' low-level events happening in between. Hence, the approaches \cite{wang2004anomaly, chakrabarti2004autopart} that identify individual events that confer a given system state are  
 inappropriate to detect sequences of such interactions between different events. Therefore, there exists a vital need for the methods that can detect the sequences of events that are related to the abnormal activities in an efficient and accurate way.

To address the aforementioned challenges, in this paper, we introduce \gid, a \underline{g}raph-based \underline{i}ntrusion \underline{d}etection algorithm to capture the interaction behavior among different entities, which sheds important light on event sequences (or sequence patterns) that are related to anomalous behavior. A key proposal to achieve this is a compact graph representation that preserves all useful information from the massive heterogeneous system monitoring data. Then, a random-walk based algorithm on this graph is conducted to provide {\em anomaly scores} that quantify the ``rareness'' or anomaly degree of the event sequences with regards to the historical data. Existing random-walk based anomaly detection algorithms (e.g., \cite{nagaraja2010p2p,liu2014isp}) can only be applied to detect single suspicious entities/events. By introducing the sender and receiver scores, we extend the random-walk technique to detect suspicious process paths, and prove its convergence in the directed acyclic graph.   


To eliminate the potential score bias from the path length, we use the Box-Cox power transformation based approach to normalize the anomaly scores so that the scores of paths of different lengths have the same distribution. 
To improve the detection accuracy, we perform the validation of the suspicious event sequences to ensure that the abnormal event sequences are sufficiently deviated from normal ones. We further propose an optimization scheme by integrating the threshold algorithm with the detection algorithm to improve the efficiency.
We launch an extensive set of experiments on a testbed that simulates the real-world attacks on the real enterprise system, to evaluate the time performance and detection accuracy of our approach. 
The results demonstrate that, on an enterprise surveillance data that contain 440 million system events, \gid\ is efficient (as much as 2 million records per minute) and accurate (higher than 80\% in terms of detection rate). We fully implement our \gid\ system and deploy it into a real enterprise security system, which greatly helps detect advanced threats, and optimize the incident response.  

To summarize, in this paper we make the following contributions:
\begin{itemize}
    \item We identify an important problem (suspicious process path discovery) in intrusion detection, and design a compact graph model to preserve all the useful information from massive system monitoring data; 
   
    \item We develop a suspicious path discovery algorithm based on random walk and Box-Cox power transformation. By introducing the sender and receiver scores, we extend the random-walk technique to detect suspicious sequences in the directed acyclic graphs, and prove its convergence; 
    \item We propose a fast optimization technique to avoid checking all candidate paths;
    
    \item We apply our algorithm to real enterprise surveillance data and demonstrate its effectiveness and efficiency.

\end{itemize}

The rest of the paper is organized as follows. Section \ref{sc:problem} formally states our problem. Section \ref{sc:approach} discusses our graph-based intrusion detection approach in detail. 
Section \ref{sc:optimization} presents the optimization scheme to improve the efficiency. 
Section \ref{sc:exp} provides the experiment results.  Section \ref{sc:related} discusses the related work. Finally, Section \ref{sc:conclusion} concludes the paper.



\section{Preliminaries}
\label{sc:problem}

In this section, we present the preliminaries of our work. For the following sections, we assume that the computer system is UNIX system, for simplicity of discussion. 

\noindent{\bf System Entities.} 
System monitoring data, collected by our agent, indicates the interactions between a set of system entities. We consider four different types of system entities: (1) {\em files}, (2) {\em processes}, (3) {\em Unix domain sockets} (UDSockets), and (4) {\em Internet sockets} (INETSockets). Each type of entities is associated with a set of attributes. 
Two types of interactions between the system entities are considered in this paper: (1) file accessed by the processes; and (2) communication between processes. 
According to the design of modern operating systems, sockets function as the proxy for different processes to communicate. 
Typically, two processes that execute on the same host communicate with each other via UDSockets, while processes on different hosts communicate with each other by INETSockets. Thus, on a single host, the interactions exist between the following types of entities: (1) processes and files, (2) processes and processes, (3) processes and sockets (both UDSockets and INETSockets), and (4) UDSockets and UDSockets. 

\noindent{\bf System Events.} 
We model the interactions between entities as system events.
Formally, a {\em system event} $e (n_b, n_d, t)$ is a record contains source entity $n_b$, destination entity $n_d$, a timestamp $t$ when $e$ happens. $n_b$ and $n_d$ are different entities from possibly different types. In computer systems, a heterogeneous event is a record involving entities of different types such as the files, processes, Unix domain sockets (UDSockets), and Internet sockets (INETSockets). And the following types of entity interactions are allowed in UNIX: (1) processes and files, (2) processes and processes, (3) processes and sockets (both UDSockets and INETSockets), and (4) UDSockets and UDSockets.  

System events can be generated at a high frequency (e.g., $20,000$ events per second). In a modern computer system, plentiful system events can happen without users' awareness. For instance, the {\em Exim} process, which is the mail transfer agent, frequently accesses the {\em /etc/hosts} file to check the mapping between host names and IP addresses. 


\noindent{\bf Event Sequence.} System events often happen in a chain. For instance, process $A$ first opens file $F$, then reads $F$, and sends the content of $F$ to process $B$. We formulate such chain as the {\em event sequence}. Formally, a sequence of events $e_1$, \dots, $e_{\ell-1}$ that happens in a chain is denoted as $S \{e_1, e_2, \dots, e_{\ell-1}\}$, where $e_i.n_d=e_{i+1}.n_b$, i.e., $e_i$'s destination entity is the $e_{i+1}$'s source entity, and $e_i.t < e_{i+1}.t$ for $i\in \{1, \dots, \ell-1\}$. The length of $S$ is $\ell$. 
The timespan of a event sequence is $ts=e_{\ell-1}.t-e_1.t$.
A simplified representation is to denote the event sequence as $S\{n_1, n_2, \dots, n_{\ell-1}, n_{\ell}\}$, informing that the data is transmitted from $n_1$ to $n_{\ell}$, via $n_2, \dots, n_{\ell-1}$ following the time order. 

\noindent{\bf Abnormal Event Sequence.} 
Most of the abnormal system behaviors, such as cyber intrusion, spying, and information stealing, often exploit event sequences to achieve their goals. Typically these event sequences, not individual events, behave differently from the regular event sequences (see formal definition of {\em abnormality} in Section \ref{sc:randomwalk}). We call those event sequences that are associated with malicious behaviors as {\em abnormal event sequences}. 
Abnormal system activities, such as cyber attacks, often finish in a relatively short time period, in order to avoid detection. Thus, the timespan of abnormal event sequences is short as well.

\nop{
Example \ref{exp:abnormal} shows an example of the abnormal event sequence.
\begin{example}
\label{exp:abnormal}
{\bf
Consider an event sequence $S$\{(``/var/log/\\install.log", ``/usr/sbin/cron", 21:45:03), (``/usr/sbin/cron", ``155.246.209.59$\rightarrow$74.125.29.156", 21:45:05)\} (or \{``/var/log/\\install.log", ``/usr/sbin/cron", ``155.246.209.59$\rightarrow$74.125.29.156"\}),
}
in which ``/var/log/install.log'' is a file, ``/usr/sbin/cron'' is a job scheduler process that normally deals with connecting to the Internet and downloading emails at regular intervals, and ``155.246.209.59$\rightarrow$74.125.29.156'' is an INETSocket. This sequence sends the data located in the {\em install.log} file to an external host (74.125.29.156) via the {\em cron} process and an INETSocket. Since the {\em install.log} file stores all the sensitive historical installation information, 
the event sequence $S$ is considered as an {\em abnormal} event sequence as it sends the sensitive file to an external host.
\end{example}
}
To measure the degree of severity of suspicious event sequences, we assign a {\em score} to each event sequence (the formulation of anomaly score is in Section \ref{sc:randomwalk}).

Intrusion attacks usually complete in a short time period, in order to avoid getting captured. In other words, only those events that are temporally close may be utilized in a sequence by the attacker to commit a cyber crime. Therefore, we mainly focus on the events in a short time span, instead of search all the historical events. More formally, our problem can be defined as follows:

\noindent{\bf Problem Statement.} 
 Given the heterogeneous event data that contains a set of events $\cal{E}$, the user-specified positive integers $\ell$, $k$, and time window size $\Delta t$, 
 we aim to find the top $k$ abnormal event sequences in $\cal{E}$ that include at most $\ell$ system events occurring within the time period of $\Delta t$.
 
 There are two major challenges in this problem: (1) How to define and compute the anomaly score of event sequence containing multiple heterogeneous entities; and (2) How to rank the event sequences of different lengths at the same time.  

\section{Algorithm}
\label{sc:approach}



\subsection{Overview}
\begin{figure}[htbp]
	\includegraphics[width=0.45\textwidth]{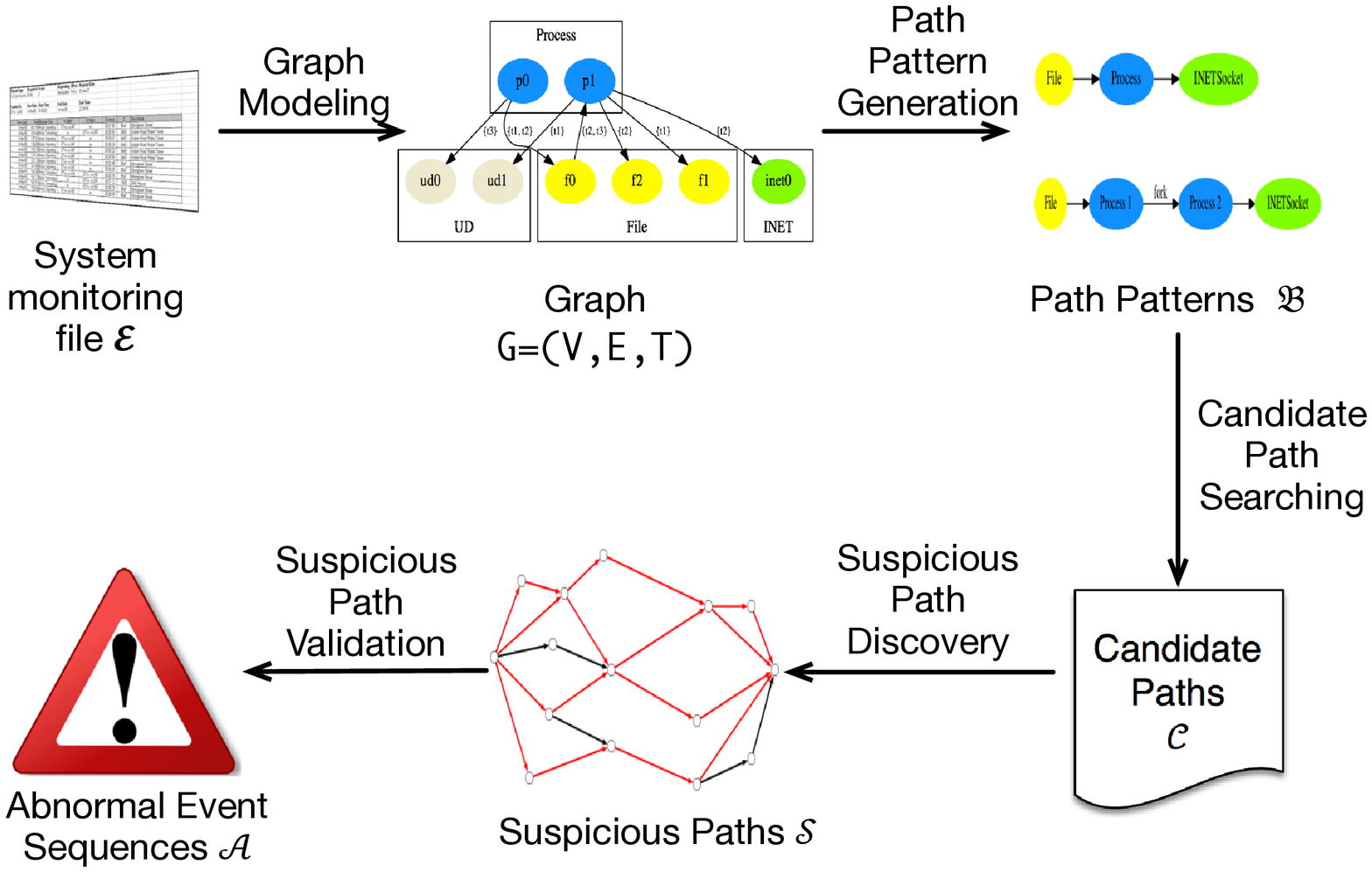}
	\vspace{-0.1in}
	\centering\caption{Framework of \gid}
	\label{fig:framework}
	\vspace{-0.2in}
\end{figure}

To address the aforementioned challenges, in this paper, we propose \gid, a \underline{g}raph-based \underline{i}ntrusion \underline{d}etection system, that can find abnormal event sequences from a large number of heterogeneous event traces. 
Figure \ref{fig:framework} shows the framework of \gid. In particular, the {\em graph modeling} component (Section \ref{sc:graph}) generates a compact graph that captures the complex interactions among event entities, aiming to reduce the computational cost in subsequent analysis components. 
The {\em path pattern generation} component 
constructs the patterns of abnormal event sequences (Section \ref{sc:pattern}). 
The {\em candidate path searching} component discovers all the candidate event sequences that comply with the path patterns by scanning the graph (Section \ref{sc:candidate}).
From the candidate paths identified, the {\em suspicious path discovery} component discriminates those abnormal event sequences from those normal ones (Section \ref{sc:randomwalk}). The distinction between abnormal and normal paths is based on the {\em anomaly scores} that measure the ``rareness'' of each candidate event sequence compared with the historical ones. 
The {\em suspicious path discovery} component returns those paths of top-k anomaly scores as suspicious paths. 
To further reduce false alarms, the {\em suspicious path validation} component measures the deviation between the suspicious sequences from the normal ones, and outputs those sequences as abnormal only if their deviation is sufficiently large (Section \ref{sc:validation}). In the following sections, we discuss the details of each component. 

\vspace{-0.15in}

\subsection{Graph Modeling}
\label{sc:graph}
The system monitoring data can be massive. For example, the data collected from a single computer system by monitoring the process interactions in one hour can easily reach 1 GB. Searching over such massive data is prohibitively expensive in terms of both time and space. Therefore, we devise a compact, graph-based representation that can greatly improve the performance of the detection procedure.


The idea of compact graph design comes from our observation that the original system surveillance data is often largely redundant in several ways. First, the redundancy comes from the attributes, as each event record always contains not only the involved entities but also the attributes of these entities. Repeatedly storing the attributes of those entities in a large number of events introduce significant redundancy. 
Second, the redundancy comes from the events as the events that involve the same entities is always repeatedly saved (with different time stamps). 
Third, normally, the abnormal behaviors, such as intrusion attacks, complete in a short time window. Therefore, it is not necessary to search the data outside of the user-defined time window.  

\vspace{-0.1in}
\begin{figure}[h]
\begin{center}
	\includegraphics[width=0.49 \textwidth]{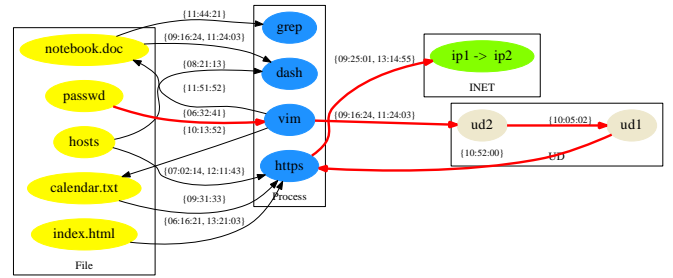}
	\vspace{-0.2in}
	\caption{\label{fig:graph}An example of compact graph model; the red path corresponds to an abnormal event sequence}
\end{center}
\vspace{-0.2in}
\end{figure}

Our graph model eliminates redundancies in the data. Given the data in a time window, we construct a directed graph $G=(V, E, T)$, with: (1) $V$ as a set of vertices, each representing an entity. For enterprise surveillance data (see Section~\ref{sec:exp:setup}), each vertex of $V$ belongs to any of the following four types: files ($F$), processes ($P$), UDSockets ($U$), and INETSockets ($I$), namely $V=F\cup P \cup U \cup I$; (2) $E$ as a set of edges. For each pair of entities $(n_i, n_j)$, if there exists any system event between them, we construct an edge $(v_i, v_j)$ in the graph, where $v_i$ ($v_j$) corresponds to $n_i$ ($n_j$); and (3) $T$ as a set of time stamps. For any edge $(v_i, v_j)$, it is possible that it is associated with multiple timestamps (i.e., the corresponding event happens multiple times). We use $T(v_i, v_j)$ to denote the set of time stamps on which this event has ever happened. Formally, $T(v_i, v_j)=\{e.t|e\in \mathcal{E}, v_i=e.n_b \quad \textrm{and} \quad v_j=e.n_d\}$.  $n_b$ ($n_d$) is the source (destination) entity of $e$. 
Given an event sequence of length $\ell$, there is a corresponding path in $G$ that includes $\ell$ vertices. In the rest of the paper, we will use event sequence and path interchangeably. 

In this paper, we are interested in only those event sequences that happen within a given time interval, 
namely the timespan is less than or equal to $\Delta t$, 
where $\Delta t$ is a given threshold. In other words, the graph only records events that happen in a time window of size $\Delta t$.  
Figure \ref{fig:graph} shows an example of the compact graph for enterprise surveillance data. 
Note that according to the types of entity interactions that are allowed in UNIX (Section \ref{sec:exp:setup}), $G$ is not a complete graph. 
Instead, it only allows the edges between (1) process and file nodes, (2) process and process nodes, (3) process and socket  (both UDSockets and INETSockets) nodes, and (4) UDSocket and UDSocket nodes. 
 
By removing the redundancy of attributes and events, our graph representation can significantly compress the original heterogeneous event data while preserving relevant information for intrusion detection. Our experiment results in Section~\ref{sec:exp:compact} demonstrate that the graph model reduces the space cost significantly. 






\subsection{Path Pattern Generation}
\label{sc:pattern}
The graph constructed by the {\em graph modeling} component can be densely connected. Path search in such graphs can be time costly. To speed up the path searching procedure, \gid\ allows users to embed the predefined set of {\em valid} path patterns $\cal{B}$ first. 

The path patterns could be defined by the experts according to their experiences and knowledge. Arguably, it would be valuable to incorporate experts' knowledge into the path pattern generation. 
Formally, given a graph $G (V, E, T)$, a path pattern $B$ is of the format $\{X_1, \dots, X_{\ell})$, where each $X_i (1\leq i\leq\ell)$ is either a specific entity (e.g., file {\em install.log}) in $V$ or a specific entity type (e.g., $P$, which can be mapped to any system process). The {\em length} of $B$ that consists of $\ell$ nodes is $\ell$. Take the information leakage problem in computer systems for example, we set each $X_i (1\leq i\leq\ell)$ as a specific system entity type, i.e., $X_i \in \{F, P, U, I\}$. Note that for all the paths that correspond to information leakage, they must satisfy that $X_1=F$ and $X_{\ell}=I$.
 
Given a path $p\in G$ and a path pattern $B$ of $G$, let $p[i]$ and $B[i]$ represent the $i$-th node in $p$ and $B$ respectively, we say $p$ is {\em consistent with} $B$, denoted as $p \prec B$, if: (1) $p$ and $B$ have the same length; and (2) for each $i$, $p[i] \in B[i]$ (i.e., the specific entity $p[i]$ belongs to the entity type $B[i]$), we say $B$ is a {\em valid} path pattern if there exists at least one path $p$ in $G$ s.t. that $p \prec B$. 
For example, the path patterns of length 3 with the four entity types for information leakage problem are $\{F, F, I\}$, $\{F, P, I\}$, $\{F, U, I\}$ and $\{F, I, I\}$. But the only valid one is $\{F, P, I\}$, because only a process node can connect a file node to a INETSocket node. 
In this way, we can extract all the valid patterns from $\mathcal{B}$ by searching in $G$.

\subsection{Candidate Path Searching}
\label{sc:candidate}
Based on the valid path patterns $\cal{B}$, 
the {\em candidate path searching} component searches for the paths in $G$ that are consistent with $\cal{B}$. Formally, 
given a set of path patterns $\cal{B}$, the {\em candidate path searching} component aims to find the set of candidate paths $\mathcal{C}$: 
\begin{equation}
\mathcal{C} = \{p|p \in G, \exists B \in \mathcal{B} \mbox{ }s.t. \mbox{ } p \prec B\} \nonumber.
\label{eq:candidate}
\end{equation}

Besides the consistency requirement, we also impose the following {\em time order constraint} on the search procedure, demanding that for each path that is consistent with $\cal{B}$, its corresponding event sequence must follow the time order.
Formally, a path $p=\{n_1, \dots, n_{r+1}\}$ satisfies the {\em time order constraint} if $\forall i \in [1, r-1]$, there exists $t_1 \in T(n_i, n_{i+1})$ and $t_2 \in T(n_{i+1}, n_{i+2})$ such that $t_1 \leq t_2$. This condition enforces the time order in the corresponding event sequences.

A straightforward approach to find all candidate paths is to apply the path patterns and time order constraints to the {\em breadth-first search} algorithm. If no valid path patterns are provided, \gid\ would only apply the time order constraints. One scan of the system event graph $G$ is sufficient to find all candidate paths. We omit the details here due to the limited space.
 
\subsection{Suspicious Path Discovery}
\label{sc:randomwalk}

It is possible that some candidate paths discovered by the {\em candidate path searching} component are not  associated with abnormal event sequences. Hence it is necessary to distinguish the suspicious paths that are highly likely to be associated with abnormal event sequences among a large set of candidate paths.

A straightforward idea of the suspicious path discovery is to define their anomaly based on the frequency of the system entities that are involved. Those paths that involve rarely-used system  entities are considered as suspicious. This is not correct as many intrusion attacks indeed only involve system entities that are popularly used in many events. 
Consider the enterprise surveillance graph in Figure \ref{fig:graph} as an example. The red path shows a typical insider attack, via which the secret {\em passwd} file is leaked through the  {\em vim} and {\em httpsd} entities. Apparently {\em vim} is the editor process and the {\em httpsd} process is a background daemon process that supports https service. Both entities are involved in many normal system events. The frequency-based anomaly approach cannot catch such intrusion attacks. 


Continuing the example, we notice that, however, the interaction between {\em vim} and {\em passwd} entities is abnormal, as typically the {\em passwd} file is accessed by the processes such as {\em bash}, but not by the {\em vim} process, which mainly serves as a file editor. 

Therefore, our basic idea is to define the anomaly based on both the system entities and the interactions among them. Each path is assigned an {\em anomaly score} that quantifies the degree of anomaly. Next, we discuss how to calculate the anomaly scores.  

First, we assign each system entity two scores, namely, a {\em sender} score and a {\em receiver} score. The sender (receiver, resp.) score measures the amounts of information that is sent (received, resp.) by the entity. 
For instance, the {\em /etc/\\passwd} file has a high sender score but relatively low receiver score, as it is sent to many processes for access permission check, but it is rarely modified. In contrast, the {\em /var/log/install.log} file has a high receiver score. 

Both sender and receiver scores are computed by performing {\em random walk} on the system event graph $G$. 
In particular, given the graph $G$, we produce a $N*N$ square transition matrix $A$, where $N$ is the total number of entities, and

\begin{equation}
A[i][j]=prob(v_i \rightarrow v_j)=\frac{|T(v_i, v_j)|}{|\sum\limits_{k=1}^N |T(v_i, v_k)||},
\label{eq:matrix1}
\end{equation}

where $T(v_i, v_j)$ denotes the set of time stamps on which the event between $v_i$ and $v_j$ has ever happened.

Intuitively, $A[i][j]$ denotes the probability that the information flows from $v_i$ to $v_j$ in $G$. We denote $A$ as
\begin{equation}
A = \left| \begin{array}{ccccc}
 & P & F & I & U\\
P & 0 & A^{P\rightarrow F} & A^{P \rightarrow I} & A^{P \rightarrow U} \\
F & A^{F\rightarrow P} & 0 & 0 & 0 \\
I & A^{I \rightarrow P} & 0 & 0 & 0 \\
U & A^{U \rightarrow P} & 0 & 0 & A^{U \rightarrow U}
\end{array} \right|,
\label{eq:matrix2}
\end{equation}
where $0$ represents a zero sub-matrix. Note that the non-zero sub-matrices of $A$ (Equation \ref{eq:matrix2}) only appear between processes and files, processes and sockets, as well as UDSockets and UDSockets, but not between processes, because the interaction between process and process does not come with information flow. This is what is allowed by the Unix system. 

The calculation of sender and receiver scores is adapted from the reasoning of authorities and hubs \cite{kleinberg1999authoritative}. In particular, 
let $x$ be the sender score vector, with $x(v_i)$ denoting the node $v_i$'s sender score. Similarly, we use $y$ to denote the receiver score vector. To calculate each node (entity)'s sender and receiver scores, first, we assign initial scores. We randomly generate the initial vector $x_0$ and $y_0$ and iteratively update the two vectors by the following
\begin{equation}
\left\{ \begin{array}{c}
x_{m+1}^T = A * y_m^T \\
y_{m+1}^T = A^T * x_m^T
\end{array}
\right.,
\label{eq:update1}
\end{equation}
where $T$ denotes the matrix transpose. According to Equation \ref{eq:update1}, an entity $v_i$'s sender score is the summation over the receiver scores of the entity to which $v_i$ sends information to. The intuition is that if an entity sends information to a large number entities of high receiver scores, this entity is an important information sender, and it should have a high sender score. Similarly, an entity should have a high receiver score if it receives information from many entities of high sender scores. 
 As a result, an entity $v_i$'s receiver score is calculated by accumulating the sender scores of the entities from which $v_i$ receives information.

From Equation \ref{eq:update1}, we derive
\begin{equation}
\left\{ \begin{array}{c}
x_{m+1}^T = (A*A^T) * x_{m-1}^T \\
y_{m+1}^T = (A^T*A) * y_{m-1}^T
\end{array}
\right. .
\label{eq:update2}
\end{equation}

In Equation \ref{eq:update2}, we update the two score vectors independently. It is easy to see that the learned scores $x_m$ and $y_m$ depend on the initial score vector $x_0$ and $y_0$. Different initial score vectors lead to different learned score values. It is difficult to choose ``good'' initial score vector in order to learn the accurate sender and receiver scores. However, we find an important property in matrix theory, namely the {\em steady state property} of the matrix \cite{jarvis1999graph}, to eliminate the effect of  $x_0$ and $y_0$ on the result scores. Specifically, let $M$ be a general square matrix, and $\pi$ be a general vector. By repeatedly updating $\pi$ with
\begin{equation}
\pi_{m+1}^T = M * \pi_{m}^T,
\label{eq:update3}
\end{equation}
there is a possible convergence state such that $\pi_{m+1}=\pi_m$ for sufficiently large $m$ value. In this case, there is only one unique $\pi_n$ which can reach the convergence state, i.e.,
\begin{equation}
\pi_n^T = M * \pi_n^T.
\label{eq:update4}
\end{equation}
The convergence state has a good property that the converged vector is only dependent on the matrix $M$, but independent from the initial vector value $\pi_0$. Based on this property, we prefer that the sender and receiver vectors can reach the convergence state. Next, we discuss how to ensure the convergence. 

To reach the convergence state, the matrix $M$ must satisfy two conditions: {\em irreducibility} and {\em aperiodicity} \cite{jarvis1999graph}. A graph $G$ is {\em irreducible} if and only if for any two nodes $v_i$, $v_j\in V$, there exists at least one path from $v_i$ to $v_j$. The period of a node $v\in V$ is the minimum path length from $v$ to $v$. The graph's period is the greatest common divisor of all the node's period value. A graph $G$ is {\em aperiodic} if and only if it is irreducible and the period of $G$ is $1$. 

As our system event graph $G$ is not always a strongly connected graph, the iteration in Equation (\ref{eq:update2}) can not reach the convergence state. To ensure convergence, we add a {\em restart matrix} $R$, which is widely used in random walk on homogeneous graph \cite{page1999pagerank} and bipartite graph \cite{sun2005neighborhood}. Typically, $R$ is a $N*N$ square matrix, with each cell value be $\frac{1}{N}$. With $R$, we get a new transition matrix $\bar{A}$:
\begin{equation}
\bar{A} = (1-c)*A + c*R,
\label{eq:restart}
\end{equation}
where $c$ is a value between 0 and 1. We call $c$ the {\em restart ratio}. With the restart technique, $\bar{A}$ is guaranteed to be an irreducible and aperiodic matrix. By replacing $A$ with $\bar{A}$ in Equation (\ref{eq:update2}), we are able to get the converged sender score vector $x$ and receiver score vector $y$. We can also control the convergence rate by controlling the restart rate value. Our experiments show that the convergence can be reached within 10 iterations.

Given a path $p=(v_1, \dots, v_{r+1})$, based on the sender and receiver score, the anomaly score is calculated as
\begin{equation}
Score(p) = 1 - NS(p),
\label{eq:anomaly}
\end{equation}
where $NS(p)$ is the regularity score of the path calculated by the following formula: 
\begin{equation}
NS(p)= \prod\limits_{i=1}^{r} x(v_i)*A(v_i, v_{i+1})*y(v_{i+1}), 
\label{eq:normaly}
\end{equation}
where $x$ and $y$ are the sender and receiver vectors, and $A$ is calculated by Equation \ref{eq:matrix2}. 
In Equation (\ref{eq:normaly}), $x(v_i)*A(v_i, v_{i+1})*y(v_{i+1})$ measures the normality of the event (edge) that $v_i$ sends information to $v_{i+1}$. Intuitively,  any path that involves at least one abnormal event is assigned a high anomaly score.
Consider the example of the suspicious path (the red path) in Figure \ref{fig:graph}. 
As the {\em passwd} file is rarely accessed by the {\em vim} process, the information transition probability between {\em passwd} and {\em vim} is low. Therefore, the event sequence is assigned with a high anomaly score.

For each path $p \in \mathcal{C}$, we calculate the anomaly score by Equation \ref{eq:anomaly}. 
However, it is easy to see that longer paths tend to have higher anomaly scores than the shorter paths. 
To eliminate the score bias from the path length, 
we {\em normalize} the anomaly scores so that the scores of paths of different lengths have the same distribution. 
Let $\mathcal{T}$ denote the normalization function. 
We use the Box-Cox power transformation function \cite{osborne2010improving} as our normalization function. In particular, let $Q(r)$ denote the set of anomaly scores of $r$-length paths before normalization. For each score $q \in Q(r)$, we apply
\begin{equation}
\mathcal{T}(q, \lambda) = \left\{
\begin{array}{lr}
\frac{q^{\lambda}-1}{\lambda} & : \lambda \neq 0 \\
\log q & : \lambda = 0
\end{array}
\right.
\label{eq:boxcox}
\end{equation}
where $\lambda$ is a normalization parameter. Different $\lambda$ values yield different transformed distributions. Our goal is to find the optimal $\lambda$ value to make the distribution after normalization as close to the normal distribution as possible (i.e., $\mathcal{T}(Q, \lambda) \sim N(\mu, \sigma^2)$). 

Next, we discuss how to compute the optimal $\lambda$. First, we assume that such $\lambda$ exists to make $\mathcal{T}(Q, \lambda) \sim N(\mu, \sigma^2)$. The density of a normalized scores is
\begin{equation}
Prob(\mathcal{T}(q, \lambda)) = \frac{\exp (-\frac{1}{2\sigma^2}(\mathcal{T}(q, \lambda)-\mu)^2)}{\sqrt{2\pi}\sigma}.
\label{eq:distribution}
\end{equation}

The profile logarithm likelihood of the normalized  distribution is
\begin{equation}
\mathcal{L}(Q, \lambda) = -\frac{n}{2}\log (\sum_{i=1}^{n} \frac{(\mathcal{T}(q_i, \lambda)-\bar{\mathcal{T}(q, \lambda)})^2}{n}) 
+ (\lambda-1)\sum_{i=1}^{n} \log q_i,
\label{eq:profile}
\end{equation}
where $\bar{\mathcal{T}(q, \lambda)}=\frac{1}{n}\sum_{i=1}^{n} T(q_i, \lambda)$.

To minimize the margin between the normalized and the normal distribution, we find the $\lambda$ that maximizes the log-likelihood. A possible solution is to take derivatives of $\mathcal{L}(Q, \lambda)$ on $\lambda$, and pick $\lambda$ that makes $\frac{\partial \mathcal{L}}{\partial \lambda} = 0$. 
The {\em suspicious path discovery} component returns those paths of top-k normalized anomaly scores as suspicious paths.

\subsection{Suspicious Path Validation}
\label{sc:validation}

To further validate the discovered suspicious paths, we calculate the {\em t-value} between the two groups of paths: 
all candidate paths $\mathcal{C}$, and the set of discovered suspicious paths $\mathcal{S}$.





 The t-test returns a confidence score that determines whether the difference between the two sets of paths is significant. If the confidence score is greater than $0.9$ with $p-value$ smaller than $0.05$, all paths in $\mathcal{S}$ are considered as abnormal paths that are relevant to intrusion attacks. Otherwise, we treat those paths as normal and do not raise alerts. 
 
 The suspicious path validation component prevents \gid from sending false alarms when there is no attack at all.
\section{Optimized Suspicious Path Discovery}
\label{sc:optimization}

The suspicious path discovery method (Section \ref{sc:randomwalk}) calculates the anomaly score for each candidate path. However, 
the number of candidate paths can be prohibitively large. 
It would be desirable if we only need to check a small number of candidate paths to find those suspicious ones. 
In this section, we devise an optimization scheme that addresses this issue by integrating the {\em threshold algorithm} \cite{fagin2003optimal} with our intrusion detection algorithm. 
The optimization scheme notably improves the efficiency of suspicious path discovery (see Section \ref{subsec:time}).


Intuitively, the top-$k$ suspicious paths are those candidate paths with the $k$ largest {\em anomaly score} $Score(p)$. 
We observe that the anomaly score function has the {\em monotone} property. In particular, given two paths $p$ and $p'$ of the same length, where $p=(v_1, \dots, v_{\ell+1})$, and $p'=(v_1', \dots, v_{\ell+1}')$, if $x(v_i)\leq x(v_i')$ and $y(v_i)\leq y(v_i')$ for $i\in [1, \ell+1]$, it must be true that $NS(p)\leq NS(p')$, thus $Score(p)\geq Score(p')$. Based on the monotone property, we design the procedure shown in Algorithm \ref{alg:optimization} to find the top-$k$ suspicious paths, without the need to calculate the anomaly score of each path. 

\begin{algorithm}
\caption{Discover top-k suspicious paths}
\label{alg:optimization}
\begin{algorithmic}
	\REQUIRE $G=(V,E,T)$, where $V=F\cup P\cup U \cup I$, $k$,
	\ENSURE $SP$ that contains the top-k suspicious paths of $G$.
	\STATE Initialize $SP$ as an empty priority queue; 
	\STATE Apply random walk on $G$ to calculate sender score vector $x$ and receiver score vector $y$; 
	\STATE Let $F_X$ be the files sorted in descendent order of the sender score. Let $F_Y$ be the files sorted in descendent order of the receiver score. 
	\STATE Create $P_X$, $P_Y$, $U_X$, $U_Y$, $S_X$ and $S_Y$ in the same way. 
	\STATE Let $E'$ be the edges sorted descendingly by $A[i][j]$
	\WHILE{$P_X$, $P_Y$, $U_X$, $U_Y$, $S_X$, $S_Y$ and $E'$ are not empty}
		\STATE $f_x=F_X.pop()$, $f_y=F_Y.pop()$
		\STATE $p_x=P_X.pop()$, $p_y=P_Y.pop()$
		\STATE $u_x=U_X.pop()$, $u_y=U_Y.pop()$
		\STATE $s_x=S_X.pop()$, $s_y=S_Y.pop()$, $e=E'.pop()$
		\STATE Assume $p$ is the path involving $f_x$, $f_y$, $p_x$, $p_y$, $u_x$, $u_y$, $s_x$, $s_y$, $e'$.
		\IF{$Score(p)\leq min\{Score(p')|p'\in SP\}$}
		    \BREAK
		\ELSE
		    \STATE Find the path set $\cal{P}$ such that every path $p$ confirms to the event sequence pattern and time constraint, and $p$ involves at least one node in $\{f_x, f_y, p_x, p_y, u_x, u_y, s_x, s_y\}$ or $e'$.
		    \FORALL{$p\in \cal{P}$}
			    \IF{$Score(p)>min\{Score(p')|p'\in SP\}$}
			        \STATE Insert $p$ into $SP$
			        \IF{$SP$ contains $k+1$ paths}
				        \STATE Remove the (k+1)-th path from $SP$.
				    \ENDIF
			    \ENDIF
		    \ENDFOR  
		\ENDIF
	\ENDWHILE
	\RETURN $SP$
\end{algorithmic}
\end{algorithm}


Our algorithm is adapted from the well-known {\em threshold algorithm} \cite{fagin2003optimal}. 
First, we apply random walk on the graph $G$ to calculate the two vectors $x$ and $y$. Second, for each type of entities, we create two queues sorted in the descendent order of the sender score and the receiver score respectively. 
Also, we sort the edges according the probability $A[i][j]$. 
After that, in each iteration of the WHILE loop, we fetch the entity or edge with the smallest score from each queue, and identify all the valid paths that contain these entities and edges. 
Assume that there is a path $p$ consisting of these entities and edges, 
we calculate $Score(p)$. Apparently $Score(p)$ is the highest anomaly score for all the paths that are not explored yet. 
If $Score(p)$ is no larger than the minimum anomaly score of all paths in the output $SP$, we stop the iterations and output $SP$. Otherwise, we discover the paths $\cal P$ that involve at least one un-checked entity that is of the highest score in any queue, and calculate the anomaly scores of these paths. Let the $k$-th path $p_k \in SP$ be the path in $SP$ that is of the minimal anomaly score. For any path $p\in\cal P$ such that $Score(p)> Score(p_{k})$, we replace $p_{k}$ with $p$. 
By Algorithm \ref{alg:optimization}, we only need to calculate the anomaly score for a small number of valid paths to find the top-k suspicious paths. It has proven that the {\em threshold algorithm} can correctly find the top k answers if the aggregation function is monotone \cite{fagin2003optimal}. Therefore, our optimization algorithm can find exact top-k suspicious  paths efficiently. 

In Algorithm \ref{alg:optimization}, we use random walk with restarts to calculate the sender and receiver scores. The complexity of the random walk step is $O(N^2)$ \cite{tong2006fast}, where $N$ is the number of entities in the graph. This is because it only needs a constant number of steps of matrix multiplication to converge. 
Suppose that there are $C$ candidate paths, the time complexity to extract the top-k suspicious ones is $O(C)$, as in the worst case, we need to calculate the anomaly score for every candidate path. Thus, the total complexity for Algorithm \ref{alg:optimization} for the worst case is $O(N^2+C)$. However, due to the early stop condition of the threshold algorithm, the average-case complexity of Algorithm \ref{alg:optimization} is $O(N^2+C')$, where $C'<< C$.

\section{Experiments}
\label{sc:exp}

\subsection{Experiment Setup}
\label{sec:exp:setup}
\noindent{\bf Data Sets.}
We use a real-world system monitoring data set in our experiments. The data is collected from an enterprise network composed of 33 UNIX machines, in a time span of three consecutive days (i.e., 72 hours). 
The sheer size of the data set is around $157$ Gigabytes. 
We consider four different types of system entities: (1) {\em files}, (2) {\em processes}, (3) {\em Unix domain sockets} (UDSockets), and (4) {\em Internet sockets} (INETSockets). Each type of entities is associated with a set of attributes and a unique identifier. 
Two types of events (i.e., interactions between the system entities) are considered in this paper: (1) file accessed by the processes; and (2) communication between processes. 
According to the design of modern operating systems, sockets function as the proxy for different processes to communicate. 
Typically, two processes that execute on the same host communicate with each other via UDSockets, while processes on different hosts communicate with each other by INETSockets. Thus, on a single host, the interactions exist between the following types of entities: (1) processes and files, (2) processes and processes, (3) processes and sockets (both UDSockets and INETSockets), and (4) UDSockets and UDSockets. 
In total, there are around {\bf 440 million} system events. These events are related to $410,166$ processes, $1,797,501$ files, $185,076$ UDSockets and $18,391$ INETSockets. 
 

\noindent{\bf Testbed and Parameters.}
We implement our algorithm in Java and run it on a PC with a 2.5GHz CPU and 8GB RAM. We set the {\em time window size} as one hour, namely we are interested in catching intrusions which occurs within an hour. We use the {\em tumbling window} model to process the stream data for simplicity. 
By default, we set the $k = 10$ and $\ell = 5$. 
Regarding the restart ratio $c$, the detection accuracy reaches a plateau as $c$ grows from $0.5$ to $0.9$, which indicates that \gid\ is insensitive to the choice of $c$ \cite{pan2004automatic}. In the experiment, we set $c=0.6$.

Based on the default setting, for each one-hour time window, \gid\ returns the top-10 most suspicious event sequences whose lengths are no larger than $5$. 


\noindent{\bf Attack Description.}
The attack testbed was built by the Russian hackers. There are $10$ different types of attacks with various lengths from 3 to 5. For each type of attacks, the hackers tried $10$ attack scenarios at different times during those three days, which results in total $300$ event sequences that correspond to intrusion attacks into the data. 
All the $10$ types of attacks exploit event sequences to transmit sensitive information to an unauthorized party. Due to the space limits, here we only list the three major types of attacks. 
\begin{itemize}
\item \noindent{\bf Type 1.} This attack targets at the {\em /selinux/\\mls} file, which defines the MLS (Multi-Level Security) classification of files within the host. In general, the {\em /selinux/mls} file should be kept secret to all users except for the security administrator, as its leakage exposes the security rules of a computer system and enables the attackers to find potential vulnerabilities. By the intrusion attack, the attacker first exploits the {\em ssh} process to access {\em /selinux/mls} file. If the file access is successful, the file content is sent to an external host (i.e., the attacker). 

\item \noindent{\bf Type 2.} This attack targets at the {\em /etc/passwd} file, which stores the password digest of all users as well as the user group information. First, the attacker tries to access the {\em /etc/passwd} file by the {\em gvfs} process, which enables easy access from a remote host via FTP. Then the attacker tries to send the file via an {\em INETSocket}. 

\item \noindent{\bf Type 3.} In this attack, the remote intruder employs the {\em bash} process to scan a secret file {\em ~/Documents/secret.xls} created by the user. We assume that the intruder has the access of the file. 
\end{itemize}

Type 1 and 2 attacks are the essential initial intrusion steps committed by the Snowden attack, while Type 3 attacks correspond to the botnet attack where the zombie computer gathers and delivers the unauthorized information to a command and control (C\&C) server.

\nop{
Considering 
there are $615,331$ normal event sequences of length 3, $28,120$ sequences of length 4, and $11,235$ sequences of length 5 in the data, 
it is a very challenging task to detect those $100$ abnormal event sequences out of total $654,786$ ones.
}

\noindent{\bf Baseline.} We compare our algorithm with two state-of-the-art algorithms: {\em NGRAM} method \cite{caselli2015sequence}, and {\em iBOAT} method \cite{chen2013iboat}. The {\em NGRAM} method has been widely studied for the identification of attacks and malicious software. This method builds the profiles of normal system behaviors from the first 4 hours, and labels those events in the following hours that do not appear in the normal profiles as abnormal ones. The {\em iBOAT} method has shown its effectiveness in suspicious trajectory discovery in GPS traces. It defines the abnormal events as those whose corresponding paths have low confidence score in the dataset. In the experiments, we set the threshold value to be $0.5$, which is already the lowest confidence we can set to ensure the highest detection rate for {\em iBOAT}.

\subsection{Detection Accuracy}
\begin{figure}[htbp]
  \begin{center}
        \includegraphics[width=0.45\textwidth]{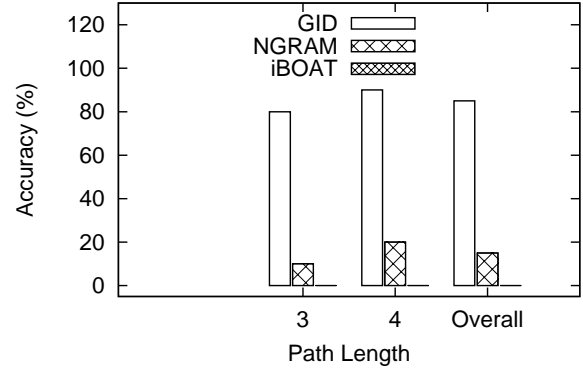}
    \caption{\small \label{fig:accuracy} Detection accuracy}
  \end{center}
\end{figure}

We also evaluate the detection accuracy of our \gid\ algorithm in catching intrusion attacks. In particular, let $h$ be the number of injected attacks, and $g$ be the number of attacks detected by \gid. The detection accuracy $a=\frac{g}{h}$. We compare the detection accuracy of \gid\ with the two baseline approaches. The results are reported in Figure \ref{fig:accuracy}. We have the following findings. First, the detection accuracy of our approach is high (with overall 85\% accuracy, and above 80\% for length-3 and length-4 attacks). The attacks \gid\ misses to catch are all type-1 attacks. The reason is that the {\em sshd} process involved in the attack has a relatively high sender score, as it frequently sends packets to other hosts. Thus the anomaly scores of the paths that involve {\em sshd} are not among the top-k list.
In comparison, the detection accuracy of the two baseline methods is much lower. In particular, the {\em NGRAM} approach is only able to detect 10\% of the length-3 attacks and 20\% of the length-4 attacks, while the {\em iBOAT} approach cannot detect any intrusion attack. We analyze the reason behind the low detection accuracy of both {\em NGRAM} and {\em iBOAT} approaches. First, most of the events in the intrusion attacks involve popular system entities such as {\em sshd} and {\em /etc/passwd}. As these system entities appear in the normal profiles, the {\em NGRAM} algorithm does not consider any path that involves these system entities as abnormal. Second, all of the simulated intrusion attacks involve an {\em INETSocket}. An {\em INETSocket} only interact with a single process in modern operating systems. This leads to a high (almost always 1) confidence score of the event sequences that correspond to the intrusion attack by the {\em iBOAT} approach. Since the {\em iBOAT} approach signals an alert if an event sequence's confidence score is below a threshold value, the {\em iBOAT} approach takes all the events that involve any INETSocket as normal sequences. 

\subsection{Time Performance}
\label{subsec:time}
\begin{figure}[htbp]
  \begin{center}
      \includegraphics[width=0.45\textwidth]{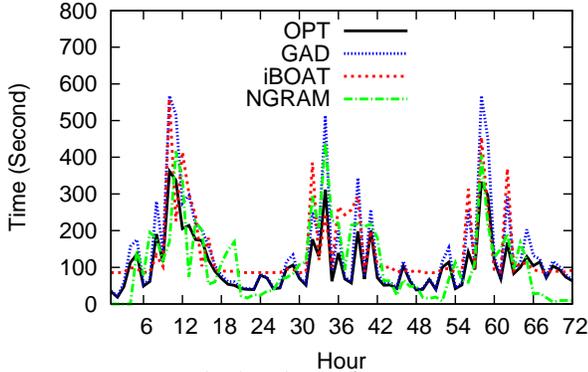}
	    \vspace{-0.2in}
    \caption{\label{fig:time} Time performance}
    \vspace{-0.25in}
  \end{center}
\end{figure}

In this section, we compare the time performance of our optimized suspicious path discovery (OPT) algorithm (see Section \ref{sc:optimization}) with the original \gid\ algorithm  (see Section \ref{sc:approach}), as well as the two baseline methods.
As varying $k$ and $\ell$ has little effect on the number of candidate paths, the time performance of \gid \ remains stable with different $k$ and $\ell$. Therefore, in Figure \ref{fig:time}, we only show the time performance when $k=10$ and $\ell=5$.
We measure the total detection time for each hour in the given 72-hour time window, except the first 4 hours (i.e., training time) for the {\em NGRAM} approach. 
In Figure \ref{fig:time}, we show the total time of the four approaches and have the following observations. First, overall, our \gid\ algorithm is very fast. It takes ten minutes at most to analyze 1-hour system events. In most cases, our detection algorithm finishes in 1 minute. The average overall detection time is $143$ seconds for \gid\ and $107$ seconds for OPT. Thus, on average, OPT saves 26\% of the total time, compared with \gid\ using exhaustive search. We also observe that when the system traffic is high and the number of candidate paths is large, the advantage of OPT is more obvious. Second, the detection time reaches its highest value at 10am of the three days, i.e., the 10-th, 34-th and 58-th hour. This is consistent with our observation that the heavy traffic of system events appears around 10am every weekday. We also compare the performance of \gid\ with the two baseline algorithms. The average execution time of the {\em NGRAM} algorithm is $121$ seconds, while the average time of the {\em iBOAT} algorithm is $144$ seconds. Given the $143$ and $107$ seconds as the average time of \gid\ and OPT, \gid\ is comparable with the two baseline methods and OPT achieves the best time performance. 
Our optimization algorithm effectively reduces the suspicious path discovery time. 



\subsection{Path Score Normalization Result}

\begin{figure}[htbp]
  \begin{center}
    \begin{tabular}{cc}
      \includegraphics[width=0.23\textwidth]{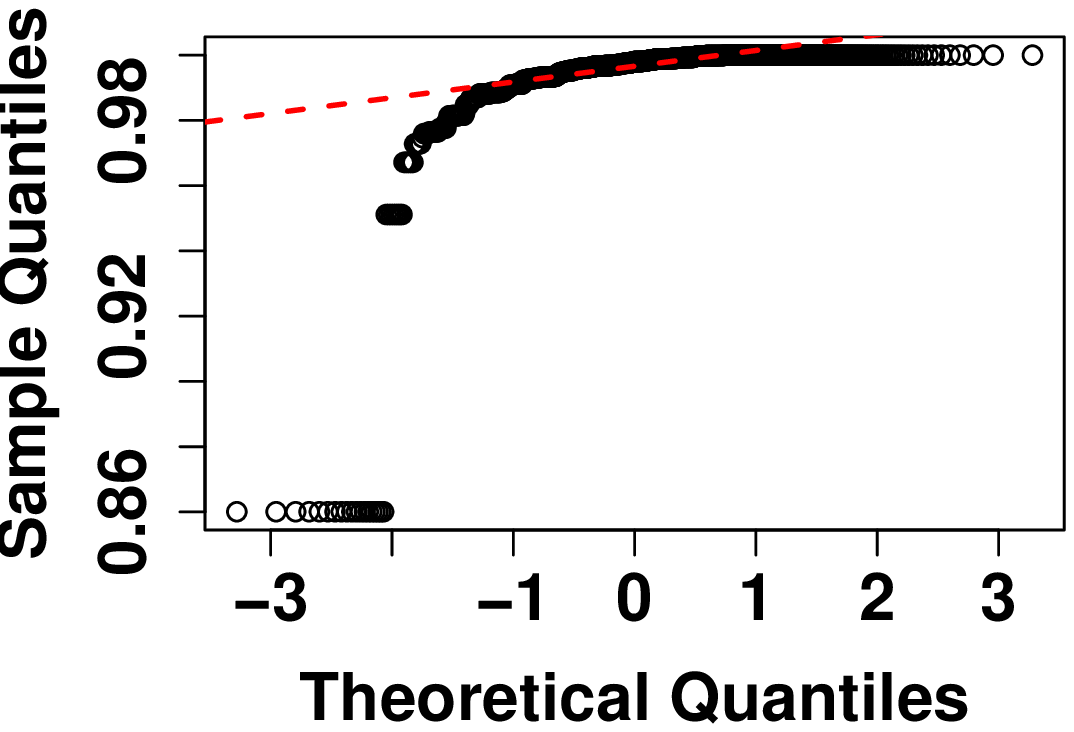}
      &
      \includegraphics[width=0.23\textwidth]{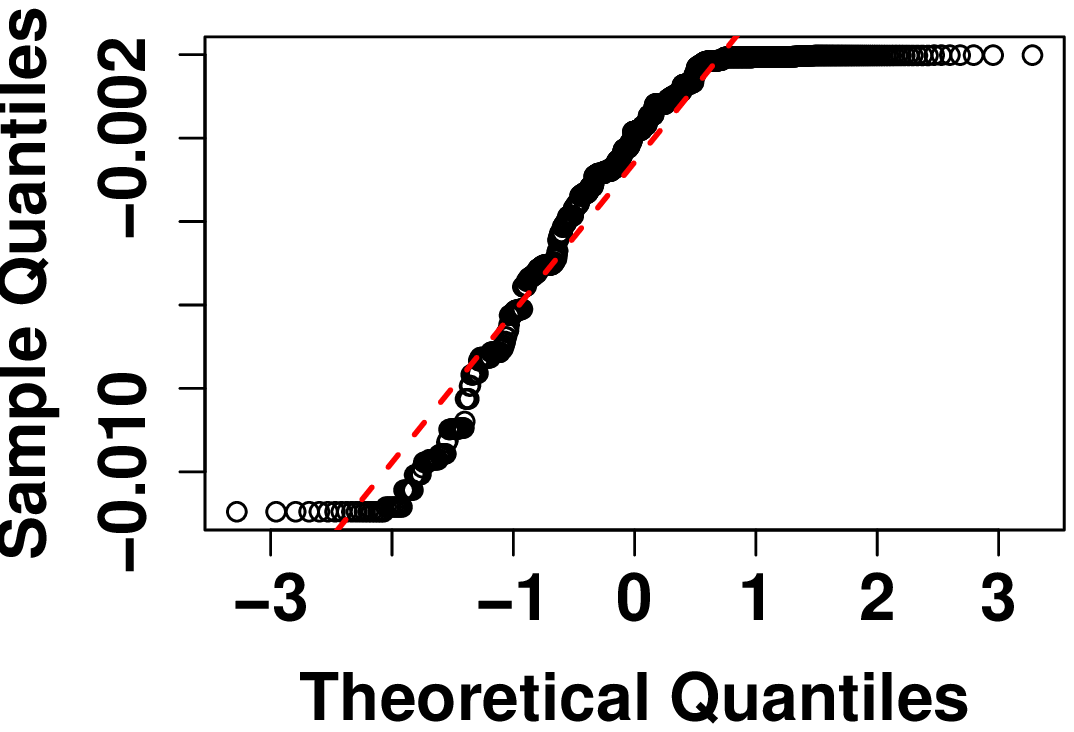}
      \\
      \small (a) Before normalization &
      \small (b) After normalization
    \end{tabular}
    \vspace{-0.2in}
    \caption{\label{fig:normalization} Normalization effect}
    \vspace{-0.25in}
  \end{center}
\end{figure}
To measure the impact of normalization to the anomaly score calculation (see Section \ref{sc:randomwalk}), we compare the anomaly scores before and after normalization. In Figure \ref{fig:normalization} (a) and (b), we show the q-q plot for the length-3 paths before and after normalization, respectively. The figure is plotted using the quantile of the anomaly score distribution against the uniform distribution. A reference line is also plotted. Intuitively, if the points fall approximately along the reference line, the score distribution is close to uniform. From Figure \ref{fig:normalization} (a), we observe a large deviation between the original anomaly score distribution and the normal distribution. After normalization, as shown in Figure \ref{fig:normalization} (b), the transformed distribution is very close to the normal distribution. 
We omit the results for length-4 and length-5 paths as they are very similar.



\subsection{Graph Compactness}
\label{sec:exp:compact}
The performance of our intrusion detection method highly relies on our graph data structure (see Section \ref{sc:graph}) . In this section, we measure the size of the constructed graph, aiming to show its compactness. 

\begin{figure}[htbp]
\begin{center}
\begin{scriptsize}
	\begin{tabular}{|c|c|c|c|c|c|}
	  \hline
	   & Process & File & INETSocket & UDSocket & Edge \\\hline
	  Avg. & 117.3 & 191.36 & 0.93 & 41.42 & 1668.4 \\\hline
	  Max & 1468 & 23290 & 130 & 6735 & 58555 \\\hline
	\end{tabular}
    \end{scriptsize}
      \\
     (a) Avg/max. number of system entities and edges 
    \\
	\includegraphics[width=0.45\textwidth]{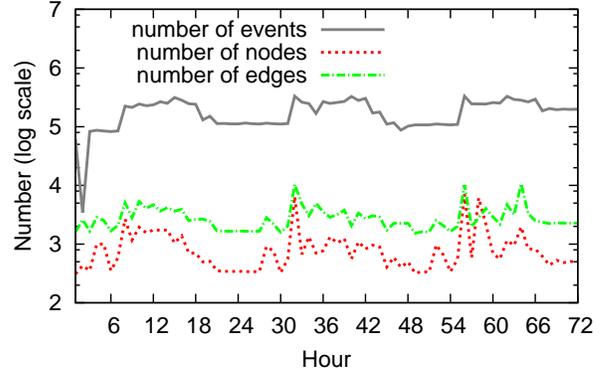}
	\\
	(b) Graph size vs. number of events
	\\
    \caption{\label{fig:memory_load} Graph compactness}
    \vspace{-0.25in}
\end{center}
\end{figure}


In Figure \ref{fig:memory_load} (a), we show the size of our system event graph in terms of number of system entities and edges. 
Specifically, we construct a graph per host per hour. Thus based on the monitoring data for 33 machines and 3 days, there are 72*33 = 2, 376 graphs. 
On average, each graph contains around $351$ nodes with four different types and less than $1.7$K edges. 
In the worst case, the graph is still within the size of $60$K edges. 
In Figure \ref{fig:memory_load} (b), we show the average size of the graphs for each hour in a 72-hour time window. The number of nodes and edges are measured by averaging the size of the 33 graphs (for 33 hosts) in one hour. 
The results show that the graph can indeed reduce the space dramatically. 
One interesting observation is that the graph reaches its largest size at the 10-th, 34-th and 58-th hour. These 3 peak hours correspond to the 10am of Day 1, Day 2 and Day 3 respectively. Normally 10am is the busiest time of system logging system as it is when most employees arrive at office.
\subsection{Distribution of Sender and Receiver Scores}

\begin{figure}[htbp]
  \begin{center}
    \begin{tabular}{cc}
      \includegraphics[width=0.23\textwidth]{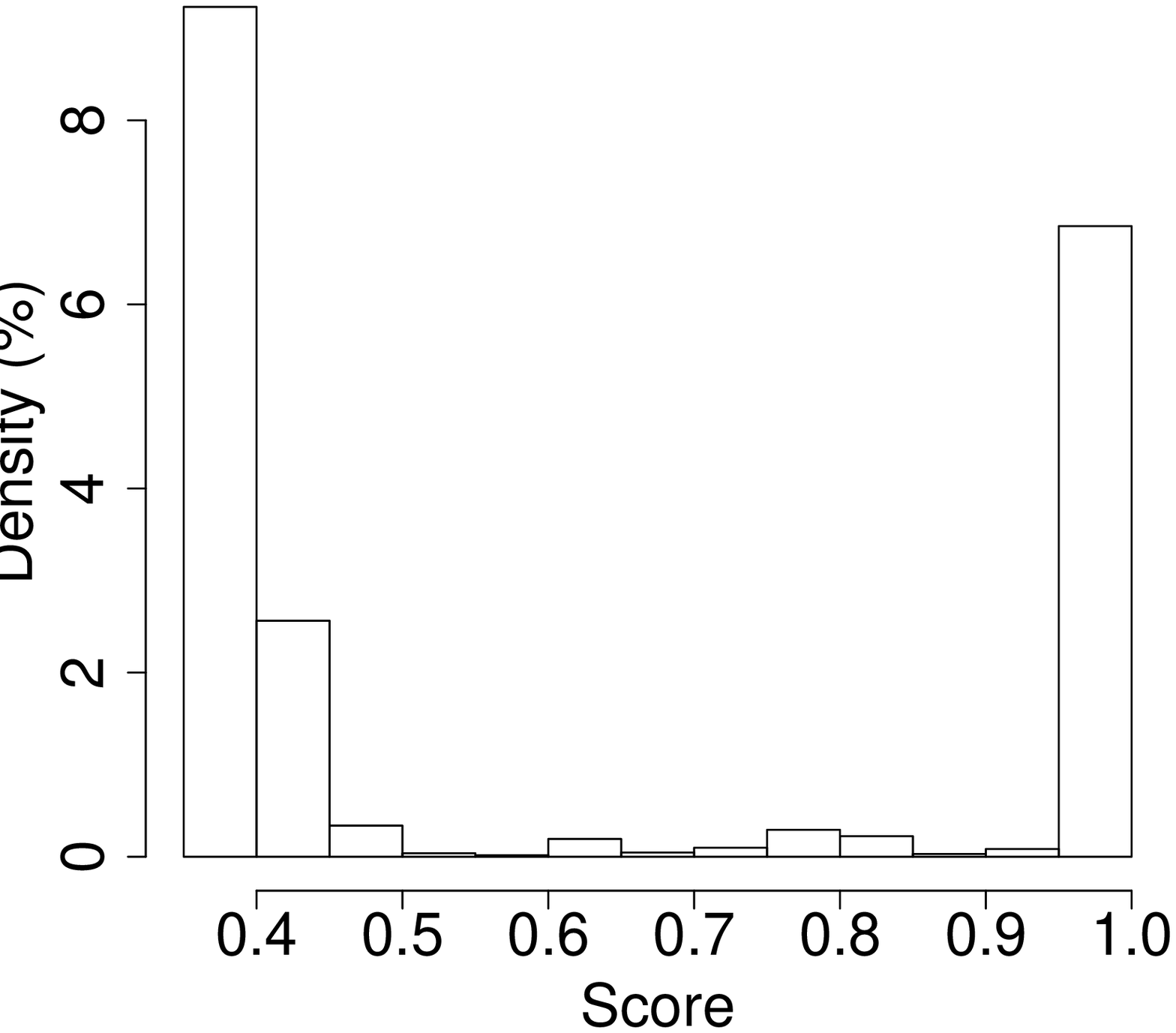}
      &
      \includegraphics[width=0.23\textwidth]{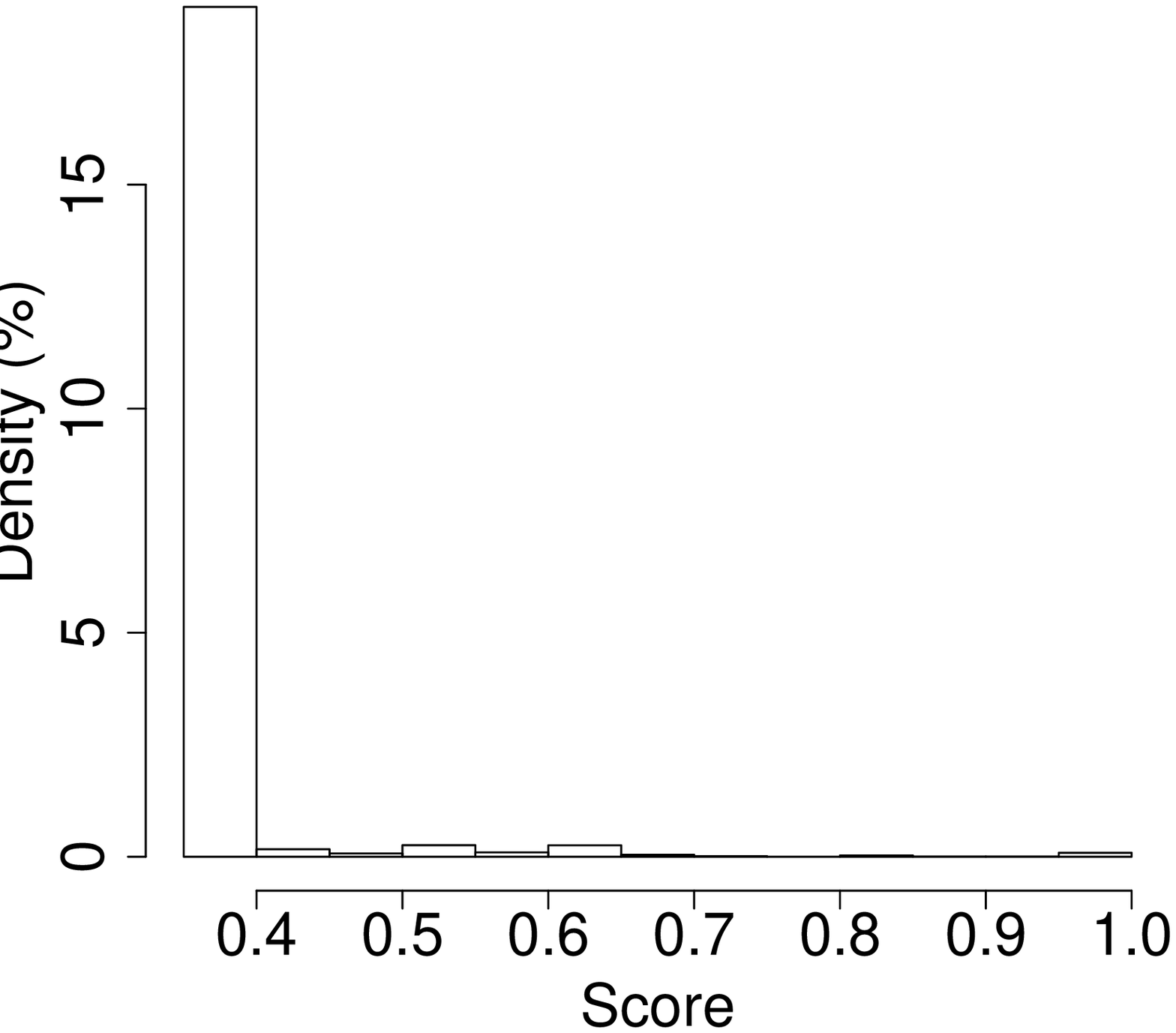}
      \\
      (a) Sender score
      &
      (b) Receiver score
    \end{tabular}
    \caption{\label{fig:random_walk} Sender and receiver score distribution}
    \vspace{-0.2in}
  \end{center}
\end{figure}

In this section, we collect the distribution of the sender and receiver scores that are computed by the random walk approach (see Section \ref{sc:randomwalk}). The purpose of showing the distribution is to demonstrate that the anomaly score, which is computed from the sender and receiver scores, indeed catches the suspicious interactions between entities. The {\em density distribution} describes the probability that the sender\\/receiver score falls into a given range. 
Figure \ref{fig:random_walk} (a) shows the density distribution of the sender scores of one specific host only, as all other hosts have similar score distribution.
It can be seen that the sender scores are of a polarized distribution. This is because many files such as shared libraries (e.g., {\em libpthread.so}) 
and system configuration files (e.g., {\em /etc/security/capability.conf}) 
are frequently accessed by various processes. Thus, the sender scores of these files are relatively high. There also exist daemon processes such as {\em /bin/dbus-daemon} and user files that never function as a information sender. This explains why a large fraction of entities have low sender scores.
Conversely, most system entities have a relatively low receiver score. This is because only a limited number of processes such as {\em cron} regularly scan files and result in high receiver scores. Meanwhile, the majority of system entities are files which are seldom updated, leading to the fact that a large portion of system entities have low receiver scores. 
Thus, our sender and receiver score computations truly models the system behaviors.

\section{Related Work}
\label{sc:related}



\subsection{Anomaly Detection}
Currently there are two types of approaches to intrusion detection, namely anomaly detection and misuse detection \cite{jones2000computer}. Anomaly detection approaches define and characterize correct/wrong behaviors of the system, while the misuse detection approaches monitor for explicit patterns, with the intrusion patterns known in advance. In this paper, we focus on anomaly detection methods. Based on the data representation, we put the existing work of anomaly detection into the following two categories.

{\bf Event-based} anomaly detection monitors and analyzes the process events of a computing system. 
Traditionally, system calls serve as a good basis for event-based analysis, as short sequences of system calls are a good discriminator for several types of intrusions \cite{caselli2015sequence}. 
\cite{caselli2015sequence} builds the profile of k-grams from normal system call traces. An alert is thrown if a new system call trace is significantly different from the normal profile. 
\cite{mutz2006anomalous} extends this work by taking the system call argument values into consideration. 
\cite{chan2003machine} considers two events with high posterior probability in the normal training data to be a good predictive pattern of normal status. In the detection phase, any violation of such pattern is recognized to be abnormal.
In all these methods, a purely normal and exhaustive set of training data is essential for constructing a robust normal profile, which substantially degrades the practicability.



In contrast, {\bf graph-based} anomaly detection represents the flow of information in a computer system with a graph and extracts abnormal substructures from it. \cite{noble2003graph, eberle2010insider} discover those small but rarely-happened substructures in the procedure of compressing the graph. The embedded {\em minimum description length} principle does not fit the highly dynamic computer system graph. To overcome this problem, \cite{sun2005neighborhood} explores the anomalies based on the community structure in an evolutionary graph. Because the concentration is limited to the graph structure, a wealth of information to describe an attack, including event timestamps and entity attributes, is disregarded. As a result, the discovered anomalies may not necessarily relate to a cyber attack.


\subsection{Random Walk}
As one of the key techniques in graph-based anomaly detection, random walk has attracted great attention because of its high effectiveness and efficiency. There are a lot of works based on the PageRank algorithm \cite{page1999pagerank}. \cite{page1999pagerank} initiates the research on random walk by using a random web surfer model to evaluate the importance of each webpage. If a random surfer stops at a webpage with high probability after a sufficiently large time, this page is of great importance. To avoid the rank sinks such as circles with no outedges, a restart matrix is taken into consideration to model the behavior that the surfer periodically gets bored and jumps to a random page. Following this work, random walk has been successfully applied to different settings such as personalized recommendation, similarity search, and information retrieval \cite{chakrabarti2007dynamic, jeh2003scaling}. But most of them focus on homogeneous graphs. Limited papers make a bold attempt to heterogeneous graphs. Among them, \cite{sun2005neighborhood} utilizes graph partition and random walk to detect anomalies in undirected bipartite graph. \cite{wieser2013multipartite} exploits the cyclic structure to rank nodes in cyclic multipartite graphs. In this paper, We extend random walk with restarts to directed acyclic graph and prove its convergence to a steady-state.

From intrusion detection perspective, recently random walk has been employed to detect single abnormal hosts/processes. For example, \cite{nagaraja2010p2p} performs random walk to identify fast-mixing components in a communication graph between Internet hosts. These components are highly likely to be P2P bots. 
\cite{liu2014isp} uses random walk to propagate the malicious scores, 
in order to discover malicious clients in a ISP network.
However, none of theses existing random walk-based algorithms can be applied to detect suspicious event sequences. By introducing sender and receiver scores and integrating with Box-Cox power transformation, we leverage random walk technique to suspicious paths discovery.

\section{Conclusion}
\label{sc:conclusion}
In this paper, we investigate the problem of detecting intrusions, especially suspicious process sequences, in enterprise system.
Different
from traditional methods that focus on detecting single entities/events, we propose \gid, a graph-based method to capture the interaction behaviour among different entities and identify abnormal event sequences. 
By leveraging random walk and Box-Cox power transformation, an event sequence is evaluated to be suspicious if any entity functions differently from its role. In this way, even the anomalous activities only involves ordinary entities, we are still able to catch such anomalies. 
The efficiency of suspicious path discovery is further improved by the proposed optimization scheme. 
Our model allows users to incorporate their domain knowledge into the path pattern generation, and also allows users to choose how many top ranked anomalies to review. We implement and deploy our approach to a real enterprise security system, and evaluate the proposed algorithm in extensive experiments. The experiment results convince us of the effectiveness and efficiency of our approach.
%

\begin{thebibliography}{10}



\bibitem{wieser2013multipartite}
N.~Becker.
\newblock Ranking on multipartite graphs.
\newblock {\em Thesis}, {\em Ludwig Maximilian University of Munich}, 2013.

\bibitem{Bellman1961}
R.~Bellman.
\newblock {\em Adaptive control processes: a guided tour}.
\newblock {\em A Rand Corporation Research Study Series}. Princeton University
  Press, 1961.


\bibitem{chakrabarti2007dynamic}
S.~Chakrabarti.
\newblock Dynamic personalized pagerank in entity-relation graphs.
\newblock{\em WWW}, pages 571--580, 2007.

\bibitem{ponemon2014cost}
L.~Phonemon.
\newblock Cost of data breach study: global analysis.
\newblock {\em Phonemon Institute Report}, 2014.


\bibitem{chen2013iboat}
C. Chen et~al.
\newblock iBOAT: Isolation-based online anomalous trajectory detection.
\newblock{\em IEEE Transaction on Intelligent Transportation Systems}, 14(2): 806-818, 2013.

\bibitem{nagaraja2010p2p}
S. Nagaraja et~al.
\newblock BotGrep: finding P2P bots with structured graph analysis.
\newblock{\em USENIX Security Symposium}, pages 95--110, 2010.

\bibitem{liu2014isp}
L. Liu et~al.
\newblock Detecting malicious clients in ISP networks using HTTP connectivity graph and flow information.
\newblock{\em ASONAM}, pages 150--157, 2014.

\bibitem{eberle2010insider}
W.~Eberle et~al.
\newblock Insider threat detection using a graph-based approach.
\newblock {\em Journal of Applied Security Research}, 6(1):32--81, 2010.

\bibitem{caselli2015sequence}
M.~Caselli et~al.
\newblock Sequence-aware intrusion detection in industrial control systems.
\newblock {\em Cyber-Physical System Security}, 
13--24, 2015.

\bibitem{jarvis1999graph}
J.~Jarvis et~al.
\newblock Graph-theoretic analysis of finite markov chains.
\newblock {\em Applied Mathematical Modeling: a Multidisciplinary Approach}, 1999.

\bibitem{jeh2003scaling}
G.~Jeh et~al.
\newblock Scaling personalized web search.
\newblock{\em WWW}, pages 271--279, 2003.

\bibitem{pan2004automatic}
J.~Pan et~al.
\newblock Automatic multimedia cross-modal correlation discovery.
\newblock{\em KDD}, pages 654--658, 2004.

\bibitem{jones2000computer}
A.~K. Jones et~al.
\newblock Computer system intrusion detection: a survey.
\newblock {\em Computer Science Technical Report}, pages 1--25, 2000.

\bibitem{jyothsna2011review}
V.~Jyothsna et~al.
\newblock A review of anomaly based intrusion detection systems.
\newblock {\em IJCA}, 28(7):26--35, 2011.

\bibitem{kleinberg1999authoritative}
J.~M. Kleinberg.
\newblock Authoritative sources in a hyperlinked environment.
\newblock {\em JACM}, 46(5):604--632, 1999.

\bibitem{lin2012intelligent}
S.-W. Lin et~al.
\newblock An intelligent algorithm with feature selection and decision rules
  applied to anomaly intrusion detection.
\newblock {\em Applied Soft Computing}, 12(10):3285--3290, 2012.

\bibitem{chan2003machine}
M.~V. Mahoney et~al.
\newblock A machine learning approach to anomaly detection.
\newblock Technical report, {\em Florida Institute of Technology}, 2003.

\bibitem{mutz2006anomalous}
D.~Mutz et~al.
\newblock Anomalous system call detection.
\newblock {\em TISSEC}, 9(1):61--93, 2006.

\bibitem{noble2003graph}
C.~C. Noble et~al.
\newblock Graph-based anomaly detection.
\newblock{\em KDD}, pages 631--636, 2003.

\bibitem{osborne2010improving}
J.~W. Osborne.
\newblock Improving your data transformations: applying the box-cox
  transformation.
\newblock {\em PARE}, 15:1--9, 2010.

\bibitem{page1999pagerank}
L.~Page et~al.
\newblock The pagerank citation ranking: bringing order to the web.
\newblock Technical report, {\em Stanford Digital Library Technologies
  Project}, 1999.


\bibitem{sun2005neighborhood}
J.~Sun et~al.
\newblock Neighborhood formation and anomaly detection in bipartite graphs.
\newblock{\em ICDM}, pages 418--425, 2005.

\bibitem{wang2004anomaly}
Y.~Wang et~al.
\newblock Anomaly intrusion detection using one class svm.
\newblock{\em Proceedings from the IEEE Information Assurance Workshop},
  pages 358--364, 2004.
  
  \bibitem{shi2012relevance}
C.~Shi et~al.
\newblock Relevance search in heterogeneous networks.
\newblock{\em EDBT}, pages 180--191, 2012.

\bibitem{sun2011pathsim}
Y.~Sun et~al.
\newblock Pathsim: meta path-based top-k similarity search in heterogeneous
  information networks.
\newblock{\em VLDB}, 2011.

\bibitem{sun2012integrating}
Y.~Sun et~al.
\newblock Integrating meta-path selection with user-guided object clustering in
  heterogeneous information networks.
\newblock{\em KDD}, pages 1348--1356, 2012.


\bibitem{chakrabarti2004autopart}
D.~Chakrabarti.
\newblock Autopart: Parameter-free graph partitioning and outlier detection.
\newblock{\em PKDD}, pages 112--124, 2004.

\bibitem{lee2012pathrank}
S.~Lee et~al.
\newblock Pathrank: a novel node ranking measure on a heterogeneous graph for
  recommender systems.
\newblock{\em CIKM}, pages 1637--1641, 2012.

\bibitem{chandola2009anomaly}
V.~Chandola et~al.
\newblock Anomaly detection: A survey.
\newblock{\em CSUR}, pages 15, 2009.

\bibitem{fagin2003optimal}
R.~Fagin et~al.
\newblock Optimal aggregation algorithms for middleware.
\newblock {\em Journal of Computer and System Sciences}, pages 614--656, 2003.

\bibitem{tong2006fast}
H.~Tong et~al.
\newblock Fast Random Walk with Restart and Its
Applications.
\newblock{\em ICDM}, pages 613-622, 2006.



\end{thebibliography}

\bibliographystyle{./IEEEtran}

\end{document}